\documentclass[10pt,aps,prl,twocolumn,superscriptaddress]{revtex4-1}

\usepackage{amsmath}
\usepackage{textcomp}
\usepackage{amssymb} 
\usepackage{graphicx}
\usepackage{tabularx}
\usepackage{placeins}
\usepackage{natbib}
\usepackage{ulem}
\usepackage[usenames]{color}

\makeatletter
\newcommand*{\rom}[1]{\expandafter\@slowromancap\romannumeral #1@} 
\makeatother

\begin{document}

\title{Shear-induced phase transition and critical exponents in 3D fiber networks}

\author{Sadjad Arzash}
\affiliation{Department of Chemical \& Biomolecular Engineering, Rice University, Houston, TX 77005}
\affiliation{Center for Theoretical Biological Physics, Rice University, Houston, TX 77030}
\author{Jordan L. Shivers}
\affiliation{Department of Chemical \& Biomolecular Engineering, Rice University, Houston, TX 77005}
\affiliation{Center for Theoretical Biological Physics, Rice University, Houston, TX 77030}
\author{Fred C.\ MacKintosh}
\affiliation{Department of Chemical \& Biomolecular Engineering, Rice University, Houston, TX 77005}
\affiliation{Center for Theoretical Biological Physics, Rice University, Houston, TX 77030}
\affiliation{Departments of Chemistry and Physics \& Astronomy, Rice University, Houston, TX 77005}

\begin{abstract}
When subject to applied strain, fiber networks exhibit nonlinear elastic stiffening. 
Recent theory and experiements have shown that this phenomenon is controlled by an underlying mechanical phase transition that is critical in nature.
Growing simulation evidence points to non-mean-field behavior for this transition and a hyperscaling relation has been proposed to relate the corresponding critical exponents. 
Here, we report simulations on two distinct network structures in 3D.
By performing finite-size scaling analysis, we test hyperscaling and identify various critical exponents.
From the apparent validity of hyperscaling, as well as the non-mean-field exponents we observe, our results suggest that the upper critical dimension for the strain-controlled phase transition is above three, in contrast to the jamming transition that represents another athermal, mechanical phase transition. 
\end{abstract}

\maketitle

Networks of interconnected fibers are common in both natural and synthetic contexts, with examples ranging from biopolymer networks to paper and carbon nanotube materials \cite{hough_viscoelasticity_2004,bryning_carbon_2007,dan_continuous_2009}. In biology, fibrous networks are primarily responsible for the mechanical stability of cells and tissues. These networks include both intracellular structures of actin and microtubules as well as extracellular matrices such as collagen and fibrin \cite{fletcher_cell_2010,pedersen_mechanobiology_2005}. In recent decades, high precision rheology experiments on purified, reconstituted biopolymer networks have revealed unusual elastic properties including negative normal stresses \cite{janmey_negative_2007} and nonlinear strain-stiffening \cite{gardel_elastic_2004,storm_nonlinear_2005}. It has been shown that the mechanics of such networks depend not only on the elasticity of individual fibers but also strongly on network connectivity. To understand the mechanical behavior of stiff biopolymer networks, coarse-grained athermal fiber models with controlled connectivity $z$ have been used in literature \cite{head_deformation_2003,head_distinct_2003,wilhelm_elasticity_2003,das_effective_2007,wyart_elasticity_2008,broedersz_modeling_2014,vermeulen_geometry_2017}. Strikingly, these simple models can accurately explain the strain stiffening observed in collagen experiments \cite{sharma_strain-controlled_2016,jansen_role_2018}. Both experiments and theory point to the importance of a mechanical phase transition as a function of strain. 

Here, we study the critical aspects of this phase transition in 3D fiber networks under applied simple shear.  Most prior systematic studies have been limited to 2D, due to the significant computational challenges imposed by the nonlinear elasticity and the need for large systems because of the diverging correlation length.  Although models to date point to qualitatively similar behavior in 2D and 3D \cite{licup_stress_2015,sharma_strain-controlled_2016,sharma_strain-driven_2016,shivers_normal_2019}, 
important questions remain, e.g., concerning the possibility of non-mean-field behavior that has been observed in 2D. 
We show that fiber networks exhibit non-mean-field behavior in 3D and are consistent with a recently identified hyperscaling scaling relation \cite{shivers_scaling_2019}, suggesting that the upper critical dimension for fiber networks is greater than three, in contrast to the jamming transition
\cite{goodrich_jamming_2014}. 

\begin{figure}[!h]
	\includegraphics[width=6cm,height=6cm,keepaspectratio]{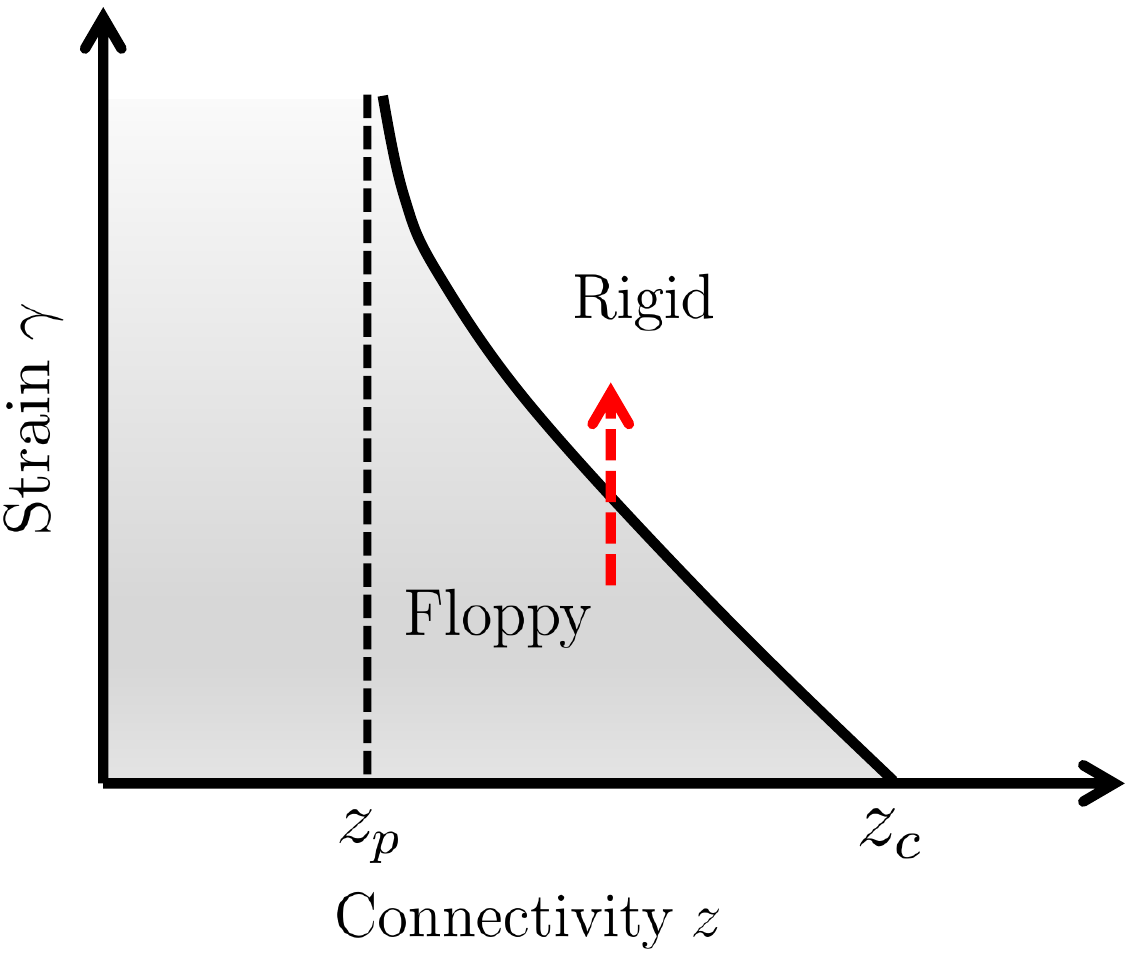}
	\caption{ \label{fig:PhaseDiagram} Schematic phase diagram for central-force networks. Networks with an average connectivity below the percolation threshold $z_p$ are disconnected, hence there is no mechanical response at any strain value. Networks with $z>z_c$, however, are stable at zero strain. In the intermediate regime, with $z_p<z<z_c$, networks are floppy at zero strain but can be rigidified by applying strain beyond a finite threshold that depends on network's connectivity and geometry. The red arrow indicates the nonlinear transition in subisostatic networks that is the subject of this study.}
\end{figure}

\textit{Model}\textemdash\, We consider two varieties of network structures; packing-derived (PD) and random geometric graph (RGG) networks at physiologically-relevant connectivity $z \lesssim 3.5$ \cite{lindstrom_biopolymer_2010,lindstrom_finite-strain_2013,jansen_role_2018}. Importantly such connectivity lies below the 3D Maxwell isostatic threshold $z_c = 6$ \cite{maxwell_i.reciprocal_1870,calladine_buckminster_1978}, at which central-force networks are marginally stable to linear order, as sketched in Fig.\ \ref{fig:PhaseDiagram}.
This connectivity-controlled rigidity transition has been extensively studied in spring networks \cite{feng_position-space_1985,arbabi_elastic_1988,wyart_elasticity_2008,broedersz_criticality_2011,feng_nonlinear_2016}, and is analogous to jamming in granular materials \cite{cates_jamming_1998,liu_jamming_1998,van_hecke_jamming_2010}, although the critical exponents differ.

For PD networks, we use a jammed packing of spheres to create an off-lattice network. 
We randomly distribute a 50-50 bidisperse mixture of spherical particles with size ratio of 1.4 in a periodic cube of size $W$, with $N = W^3$ spheres in total. The particles interact via a repulsive harmonic potential \cite{ohern_random_2002,ohern_jamming_2003,goodrich_jamming_2014}. We start by swelling the particles until the pressure becomes finite, which yields a network with $z \approxeq 2d = 6$. We then randomly remove bonds to obtain the desired connectivity $z < 6$. For RGG networks, we randomly distribute nodes in a cube with side length $W$ before connecting pairs of nodes according to a distance-dependent probability distribution \cite{beroz_physical_2017} until the desired connectivity is reached.

The network elastic energy $H$ consists of stretching and bending contributions
\begin{equation} \label{eq:NetworkEnergy}
	H = \frac{\mu}{2} \sum_{ij}^{}\frac{(\ell_{ij} - \ell_{ij,0})^2}{\ell_{ij,0}} +
	 \frac{\kappa_b}{2}\sum_{ijk}^{}\frac{(\theta_{ijk} - \theta_{ijk,0})^2 }{\ell_{ijk,0}},
\end{equation}
where $\mu$ is the stretching modulus of the individual bonds, $\ell_{ij,0}$ is the initial bond length prior to any deformation, $\ell_{ij}$ is the current bond length, $\kappa_b$ is the bending stiffness of fibers, $\theta_{ijk,0}$ is the initial angle between two adjacent bonds $ij$ and $jk$ prior to any deformation, $\theta_{ijk}$ is the current angle between adjacent bonds $ij$ and $jk$, and $\ell_{ijk,0} = \frac{\ell_{ij,0} + \ell_{jk,0}}{2}$ is the initial average bond length of bonds $ij$ and $jk$. We set $\mu = 1$ and vary the dimensionless bending stiffness $\kappa = \kappa_b/\mu \ell_c^2$, where $\ell_c$ is average initial bond length.

To study the mechanical transition, we apply a simple shear deformation $\gamma$ in a step-wise manner in the $x-z$ plane using Lees-Edwards boundary conditions \cite{lees_computer_1972}. 
Although we focus on shear, we note that uniaxial and bulk deformations can also rigidify such networks \cite{sheinman_nonlinear_2012,van_oosten_uncoupling_2016,vahabi_elasticity_2016,arzash_stress-stabilized_2019,merkel_minimal-length_2019}. 
Using the FIRE algorithm \cite{bitzek_structural_2006}, we minimize the elastic energy defined in Eq.\  \eqref{eq:NetworkEnergy} at each strain step and calculate the stress tensor as  \cite{shivers_scaling_2019}
\begin{equation} \label{eq:StressTensor}
\sigma_{\alpha \beta} = \frac{1}{2V} \sum_{ij}^{}f_{ij,\alpha} r_{ij,\beta}
\end{equation}
where $V$ is the volume of system, $f_{ij,\alpha}$ is the $\alpha$ component of the force exerted on node $i$ by node $j$, and $r_{ij,\beta}$ is the $\beta$ component of the displacement vector connecting nodes $i$ and $j$. We then compute the differential shear modulus as $K = \partial \sigma_{xz}/ \partial \gamma = \partial^2H/\partial \gamma^2$. Unless otherwise stated, quantities reported throughout the paper correspond to averages over 40 random realizations.

\textit{Scaling relation}\textemdash\, When we apply sufficiently large shear strain to a subisostatic network with central-force interactions (red arrow in Fig.\ \ref{fig:PhaseDiagram}), the system undergoes a phase transition from a floppy to a rigid state. At the critical strain $\gamma_c$, which is a function of network connectivity $z$ \cite{wyart_elasticity_2008,sharma_strain-controlled_2016}, the differential shear modulus $K$ discontinuously jumps from 0 to a finite value $K_c$ \cite{vermeulen_geometry_2017,merkel_minimal-length_2019,arzash_finite_2020}. The excess shear modulus $K - K_c$ exhibits a power law scaling behavior near $\gamma_c$, with $K - K_c \sim |\Delta \gamma|^f$, where $\Delta \gamma = \gamma - \gamma_c$. Including weak bending interactions between adjacent bonds stabilizes the network in the subcritical regime $\gamma < \gamma_c$, such that the floppy-to-rigid transition becomes a transition between bending- and stretching-dominated states. The following Widom-like \cite{widom_equation_1965} scaling function captures the mechanics of networks with finite bending stiffness \cite{sharma_strain-controlled_2016}
\begin{equation} \label{eq:Widom}
	K \approx |\Delta \gamma|^f \mathcal{G}_{\pm} \big( \frac{\kappa}{|\Delta \gamma|^\phi} \big)
\end{equation}
in which the positive and negative branches of the scaling function $\mathcal{G}_\pm$ correspond to $ \Delta \gamma > 0$ and $\Delta \gamma < 0$, respectively.  For $\gamma < \gamma_c$ and $\kappa/\Delta \gamma ^\phi \ll 1$, the shear modulus scales as $K \sim \kappa |\Delta \gamma|^{-\lambda}$ with $\lambda = \phi - f$. We note that continuity of $K$ as a function of $\gamma$ requires that $K  \sim \kappa^{f/\phi}$ when $\kappa/\Delta \gamma ^\phi$ is large.

To relate the scaling exponents near the critical strain $\gamma_c$, we follow the approach of Kadanoff \cite{kadanoff_scaling_1966} for the elastic energy per node $h$ as a function of the small parameters $t = \gamma - \gamma_c$ and $\kappa$ \cite{shivers_scaling_2019}. Rescaling the system by a factor $L$ results in a renormalized energy $h(t^\prime, \kappa^\prime) = L^d h(t,\kappa)$, in which $t^\prime$ and $\kappa^\prime$ are the renormalized variables after transformation and $d$ is dimensionality. We assume that $t^\prime = t L^{x}$ and $\kappa^\prime = \kappa L^{y}$ near the critical point, with positive $x$ and $y$. We can therefore write the elastic energy per node as \cite{shivers_scaling_2019}
\begin{equation} \label{eq:RenormalizedEnergy}
	h(t,\kappa) = L^{-d} h(t L^{x}, \kappa L^{y}).
\end{equation}
We find the differential shear modulus $K\sim L^{-d+2x}h_{2,0}(tL^x,\kappa L^y)$ from the second derivative with respect to $\gamma$ or $t$, in a way 
analogous to the heat capacity in a thermal phase transition, in which case differentiation is with respect to the temperature. Here, $h_{2,0}$ represents the second partial derivative of $h$ with respect to the first argument. So far, the rescaling factor $L$ is simply mathematical parameter that has not been specified. Thus, physical quantities such as $K$ cannot depend on $L$, from which the form of Eq.\ \eqref{eq:Widom} follows \cite{kadanoff_static_1967}.
By choosing $L = |t|^{-1/x}$, we find that $f = d/x - 2$ for $\gamma > \gamma_c$. We also identify the correlation length $\xi \sim L \sim |t|^{-\nu}$ and the scaling relations \cite{shivers_scaling_2019,kadanoff_static_1967}
\begin{equation}\label{eq:ScalingRelation}
	f = d \nu - 2 \; \mathrm{and} \; \phi = y \nu
\end{equation}
The first of these is a hyperscaling relation that is of particular importance for the appearance of the dimensionality $d$ of the system. This can can only be satisfied for mean-field systems at a particular $d=d_u$, which sets the upper critical dimension. 
Only for dimensionality below this are non-mean-field critical exponents possible.  

\begin{figure}[!h]
	\includegraphics[width=10cm,height=10cm,keepaspectratio]{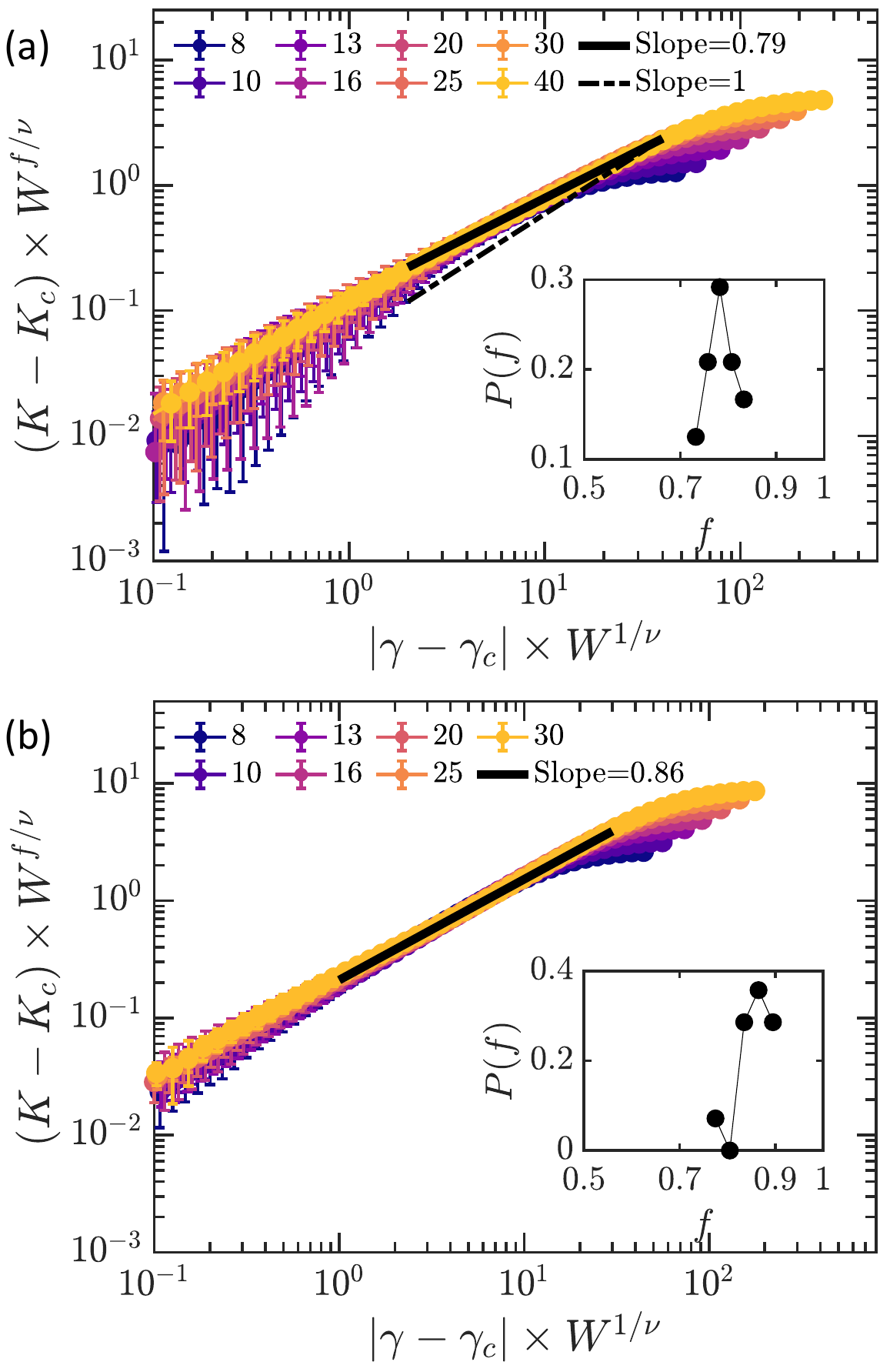}
	\caption{ \label{fig:FiniteSizeAnalysis} (a) Finite-size analysis of $K$ for the PD model at $z=3.3$ with $\kappa = 0$. In the critical region, we obtain $f = 0.79 \pm 0.03$. (b) Similar finite-size scaling analysis for PD networks at a different connectivity $z=4.0$. We find an exponent of $f = 0.86 \pm 0.04$ in the critical regime. For both models, an apparent exponent of 1.0 is observed in the finite-size dominated region, consistent with analytic behavior. The insets show the distribution of $f$ exponent.}
\end{figure}

\textit{Results}\textemdash\, In order to find the critical exponent $f$, care must be taken to account for finite size effects. 
For a finite system, only when the system size exceeds the the correlation length $\xi$, i.e., $W|t|^{\nu} \gtrsim 1$, will the thermodynamic properties approximate those of the thermodynamic limit. 
In the opposite limit $W|t|^{\nu} \lesssim 1$, which occurs close to $t=0$, correlations are limited and analytic behavior is expected. Thus, we determine the critical exponents only for $W \gtrsim \xi$. For small $t>0$, we expect the shear modulus to vary with system size $W$ as \cite{arzash_finite_2020}
\begin{equation}\label{eq:FiniteSizeCF}
	K - K_c = W^{-f/\nu} \mathcal{F}(t W^{1/\nu}).
\end{equation}
Here, $\mathcal{F}(x)$ is a scaling function that is expected to increase as $\sim x^f$ for large arguments, in order to obtain a well-defined thermodynamic limit. 
Figure\ \ref{fig:FiniteSizeAnalysis} shows the finite-size analysis corresponding to this. 
Here, the sample-dependent critical strain $\gamma_c$ is found using the bisection method \cite{merkel_minimal-length_2019,arzash_finite_2020}. For PD networks at $z=3.3$ of size $W=40$ at strains beyond the regime dominated by finite-size effects (i.e., $W \gtrsim \xi$) we find $f = 0.79 \pm 0.03$, where the errors are standard deviations for 20 random realizations. Upon increasing the connectivity to $z = 4.0$, the resulting data are
consistent with a slightly larger $f = 0.86 \pm 0.04$, which is obtained by averaging 20 samples of size $W=30$ (see Fig.\ \ref{fig:FiniteSizeAnalysis}). The distributions of the $f$ exponent are shown as insets in Fig.\ \ref{fig:FiniteSizeAnalysis}. For the RGG model at $z=3.3$ we find $f = 0.92 \pm 0.02$ using a system size of $W=30$. \footnote{We note that the RGG model, similar to lattice-based models, has long straight fibers. In this case, for small system sizes, small connected clusters of nearly aligned bonds can span the length of the simulation box. These clusters, which may comprise only a small fraction of the network's bonds, nonetheless determine the critical strain at which the $K$ becomes nonzero. As a result, in some cases, such networks exhibit an apparent two-branch behavior, with a regime of unusually low stiffness immediately above the critical strain followed by a more typical stiffening regime at larger strains. Similar behavior has been observed in 2D triangular networks \cite{arzash_finite_2020}. We demonstrate this effect for a set of RGG network samples in Fig.\ A.3 in the Appendix. To calculate $f$ for RGG networks, we removed samples exhibiting this two-branch behavior from our ensemble (see the discussion in the Appendix).}

Similar to 2D models \cite{arzash_finite_2020}, we find that the shear modulus discontinuity $K_c$ decreases as system size $W$ increases (see Fig.\ \ref{fig:A_Kc_Z33} in the Appendix). We note that as network size increases, the regime over which $f$ can be computed extends to smaller $|\Delta\gamma|$. Here, a non-mean-field exponent of $f<1$ is seen. In the finite size-dominated regime, however, the data are consistent with $f=1$, which can be explained by a leading first term in the scaling function that becomes analytic when $W \lesssim \xi$, as previously seen in 2D fiber networks \cite{arzash_finite_2020} and jammed systems \cite{goodrich_finite-size_2012}.

\begin{figure}[!h]
	\includegraphics[width=10cm,height=10cm,keepaspectratio]{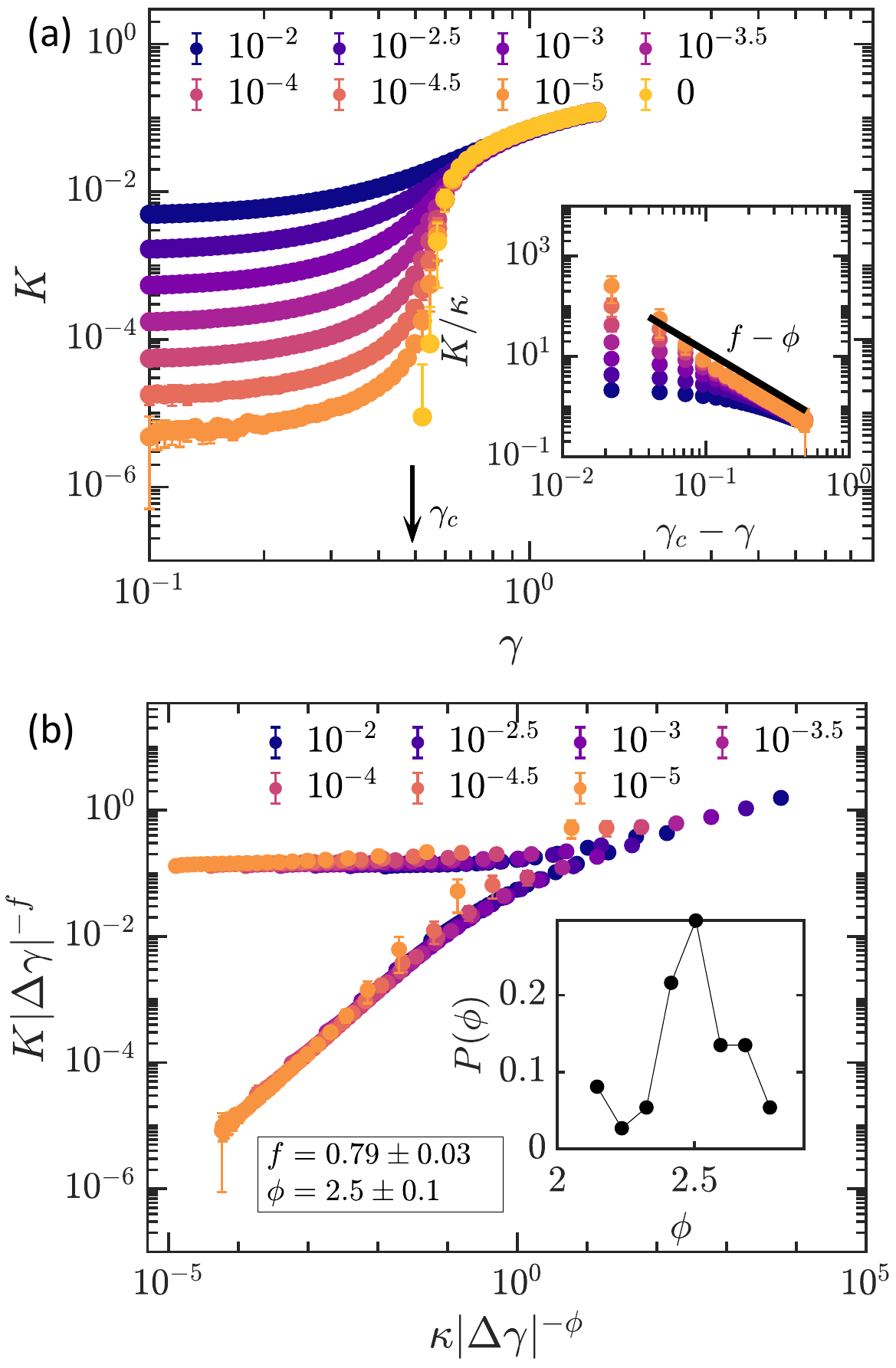}
	\caption{ \label{fig:Widom} (a) Differential shear modulus $K$ for PD networks with $z=3.3$ and system size $W=30$ at various bending stiffness $\kappa$ as shown in the legend. The inset shows the scaling behavior of $K$ in the subcritical region, where $K \sim \kappa |\Delta \gamma|^{-\lambda}$ with $\lambda = \phi - f$. (b) Widom-like scaling collapse of $K$ for data in (a). Using the critical exponents $f$ and $\phi$ as explained in the text, we are able to collapse our data based on Eq.\ \eqref{eq:Widom}. The inset shows the distribution of $\phi$ values.}
\end{figure}

In Fig.\ \ref{fig:Widom}a, we plot the shear modulus versus strain for packing-derived networks with finite $\kappa$ and $z=3.3$. From the value of $f$ above, we find the critical exponent $\phi$ in the subcritical regime $\gamma < \gamma_c$ from Eq.\ \eqref{eq:Widom}. Since $K$ must be proportional to $\kappa$ in this regime, we expect $K\sim\kappa |\Delta \gamma|^{f-\phi}$ (see inset). The distribution of $\phi$ values with $\phi = 2.5 \pm 0.1$ is shown in the inset of Fig.\ \ref{fig:Widom}b. Considering Eq.\ \eqref{eq:Widom}, we expect to find a scaling collapse of data in Fig.\ \ref{fig:Widom}a for various $\kappa$ values to a single master curve. Figure\ \ref{fig:Widom}b shows this Widom-like scaling analysis. As we can see, using the obtained values of $f$ and $\phi$ the data collapse in two branches, one for the data above $\gamma_c$, one for the data below $\gamma_c$. Figures\ \ref{fig:A_Finite_Kappa_Z33_RGG} and \ref{fig:A_Widom_Z33_RGG} in the Appendix show similar analysis for RGG model with finite bending rigidity that results in $\phi = 2.8 \pm 0.2$.

\begin{figure}[!h]
	\includegraphics[width=10cm,height=10cm,keepaspectratio]{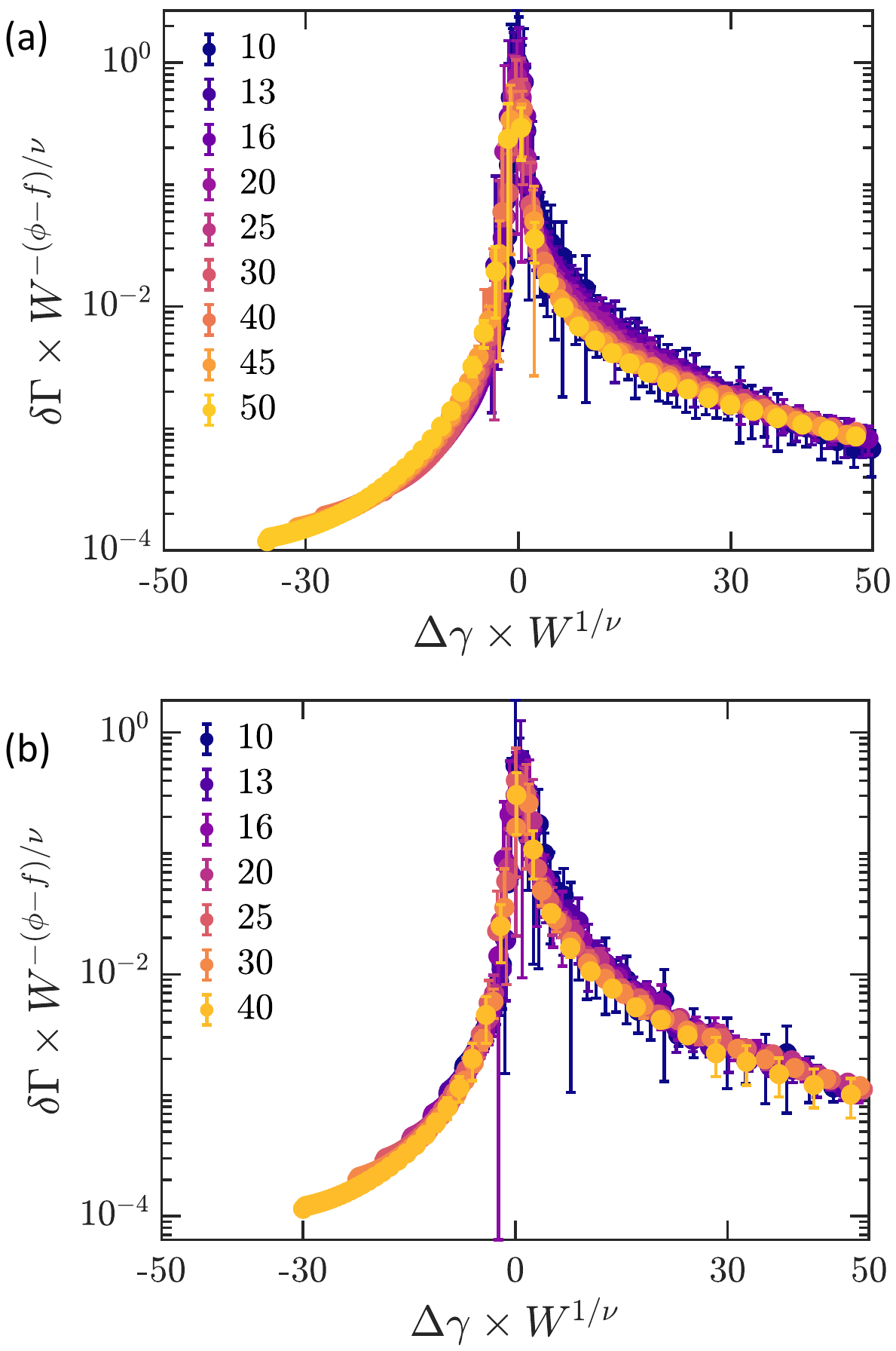}
	\caption{ \label{fig:Diffnonaffinity} (a) The finite-size analysis of the nonaffine fluctuations for central-force PD networks at $z=3.3$. As explained in the main text, the scaling exponents $f$ and $\phi$ are obtained using shear modulus data in regimes $\Delta \gamma > 0$ and $\Delta \gamma < 0$, respectively. The correlation length exponent $\nu$ is found using the scaling relation $f = d \nu -2$. (b) Similar scaling collapse of nonaffine fluctuations as in (a) for central-force RGG model at $z=3.3$.}
\end{figure}

One of the most striking features of a critical phase transition is the divergence of fluctuations at the critical point, along with the divergent correlation length $\xi$ for these fluctuations. Here, we measure the nonaffine displacement fluctuations of network nodes under an infinitesimal shear strain, defined as
\begin{equation}\label{eq:DiffNonaffinity}
\delta \Gamma = \frac{\langle || \mathbf{u} - \mathbf{u}^\mathrm{af}||^2\rangle}{\ell_c^2 \delta \gamma^2}
\end{equation}
where the $\mathbf{u}$ and $\mathbf{u}^{\mathrm{af}}$ are the relaxed and affine displacement vectors of network nodes, respectively,  after applying a small strain $\delta \gamma$, $\ell_c$ is the average initial bond length of the network, and the angular brackets represent an average over all nodes. For central-force networks, the fluctuations $\delta \Gamma$ diverge as the network approaches the critical strain \cite{sharma_strain-driven_2016}. Since the nonaffine displacements $\delta u^2$ are found by minimizing the energy $h(t,\kappa)$, we expect that $h \sim \kappa \delta u^2 \sim \kappa \delta \gamma^2 \delta \Gamma$ for small but finite $\kappa$. Therefore, the fluctuations diverge as \cite{shivers_scaling_2019, broedersz_mechanics_2011}
\begin{equation}
	\delta \Gamma \sim |\Delta \gamma|^{-\lambda}\label{eq:fluctuations}
\end{equation}
where the same $\lambda = \phi - f$ is observed for both $\gamma<\gamma_c$ and 
$\gamma>\gamma_c$.

To verify the scaling relation in Eq.\ \eqref{eq:ScalingRelation}, we calculate the nonaffine fluctuations $\delta \Gamma$ for networks with $\kappa = 0$, as shown for various system sizes in Fig.\ \ref{fig:A_DiffNonaffinity_not_scaled_Z33} in the Appendix. From Eq.\ \eqref{eq:fluctuations}, we expect the following scaling form to capture the behavior of $\delta \Gamma$ in finite simulations: \cite{sharma_strain-driven_2016}
\begin{equation}
	\delta \Gamma = W^{\lambda/\nu} \mathcal{H}(\Delta \gamma W^{1/\nu}),
\end{equation}
where $\mathcal{H}$ is a scaling function and $\lambda = \phi -f $. Figure\ \ref{fig:Diffnonaffinity}a shows the finite-size scaling collapse of $\delta \Gamma$ data using the previously obtained values of $f$ and $\phi$. The correlation length exponent $\nu$ is computed from the scaling relation in Eq.\ \eqref{eq:ScalingRelation}. Thus, this collapse demonstrates the validity of the hyperscaling relation $f = d \nu -2$ in 3D systems. A similar scaling collapse of fluctuations is shown in Fig.\ \ref{fig:Diffnonaffinity}b for the RGG model at $z=3.3$. In order to further test this, we also simulated 3D PD networks at a different connectivity $z=4.0$. The figures for this are shown in Figs.\ A.9-13 in the Appendix. The scaling exponents $f$ and $\phi$ are slightly larger than the corresponding exponents for networks with $z=3.3$. With these new scaling exponents, we perform a similar finite-size scaling analysis to Fig.\ \ref{fig:Diffnonaffinity} and find further evidence that the hyperscaling relation holds.

\textit{Conclusion}\textemdash\, In this study, we measure the exponents associated with the strain-driven rigidity transition for subisostatic 3D spring networks under applied simple shear. In agreement with previous work on various network architectures \cite{sharma_strain-controlled_2016,rens_nonlinear_2016,shivers_scaling_2019}, we find non-mean-field exponents in the critical regime by performing a systematic finite-size analysis in our 3D computational models. We also demonstrate evidence to support a recently proposed hyperscaling relation between critical exponents \cite{shivers_scaling_2019}. 
Taken together, these results point to an upper critical dimension $d_u>3$ for the strain-controlled phase transition that is above three, in stark contrast to the jamming transition \cite{goodrich_jamming_2014}. 
While our focus here has been on the subisostatic transition that is most relevant to fiber networks such as collagen, it is interesting to note that the isostatic critical point corresponding to $z=6$ in 3D has also been shown to exhibit non-mean-field exponents \cite{broedersz_criticality_2011}. 
In future work, it will be interesting to study hyperscaling for that transition as well. 

\section*{Acknowledgments}

This work was supported in part by the National Science Foundation Division of Materials Research (Grant DMR-1826623) and the National Science Foundation Center for Theoretical Biological Physics (Grant PHY-2019745). 

%

\newpage
\onecolumngrid
\appendix
\renewcommand\thefigure{A.\arabic{figure}} 
\setcounter{figure}{0}
\renewcommand\appendixname{}
\renewcommand{\theequation}{A.\arabic{equation}}
\setcounter{equation}{0}

\section*{Appendix}

\subsection*{3D PD networks at $z=3.3$}

\begin{figure}[!h]
	\includegraphics[width=10cm,height=10cm,keepaspectratio]{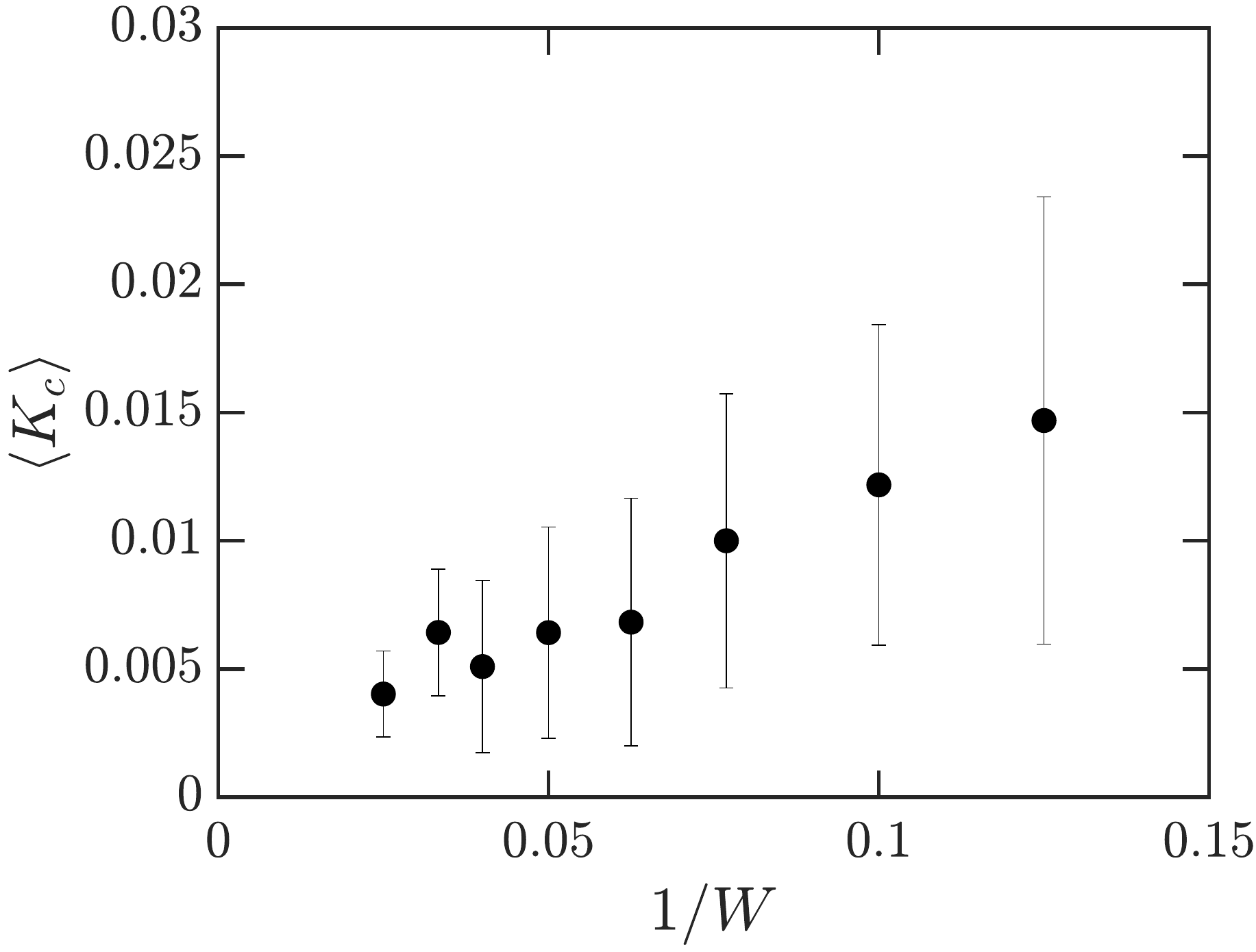}
	\caption{ \label{fig:A_Kc_Z33} The ensemble average of shear modulus discontinuity $K_c$ for 3D PD networks at $z=3.3$ versus the inverse of system size $1/W$. Similar to 2D networks, we find that $K_c$ decreases as we increase $W$. However, due to large finite-size effects, our data are not inconsistent with a vanishing $K_c$ in the thermodynamic limit.}
\end{figure}
\FloatBarrier

\begin{figure}[!h]
	\includegraphics[width=10cm,height=10cm,keepaspectratio]{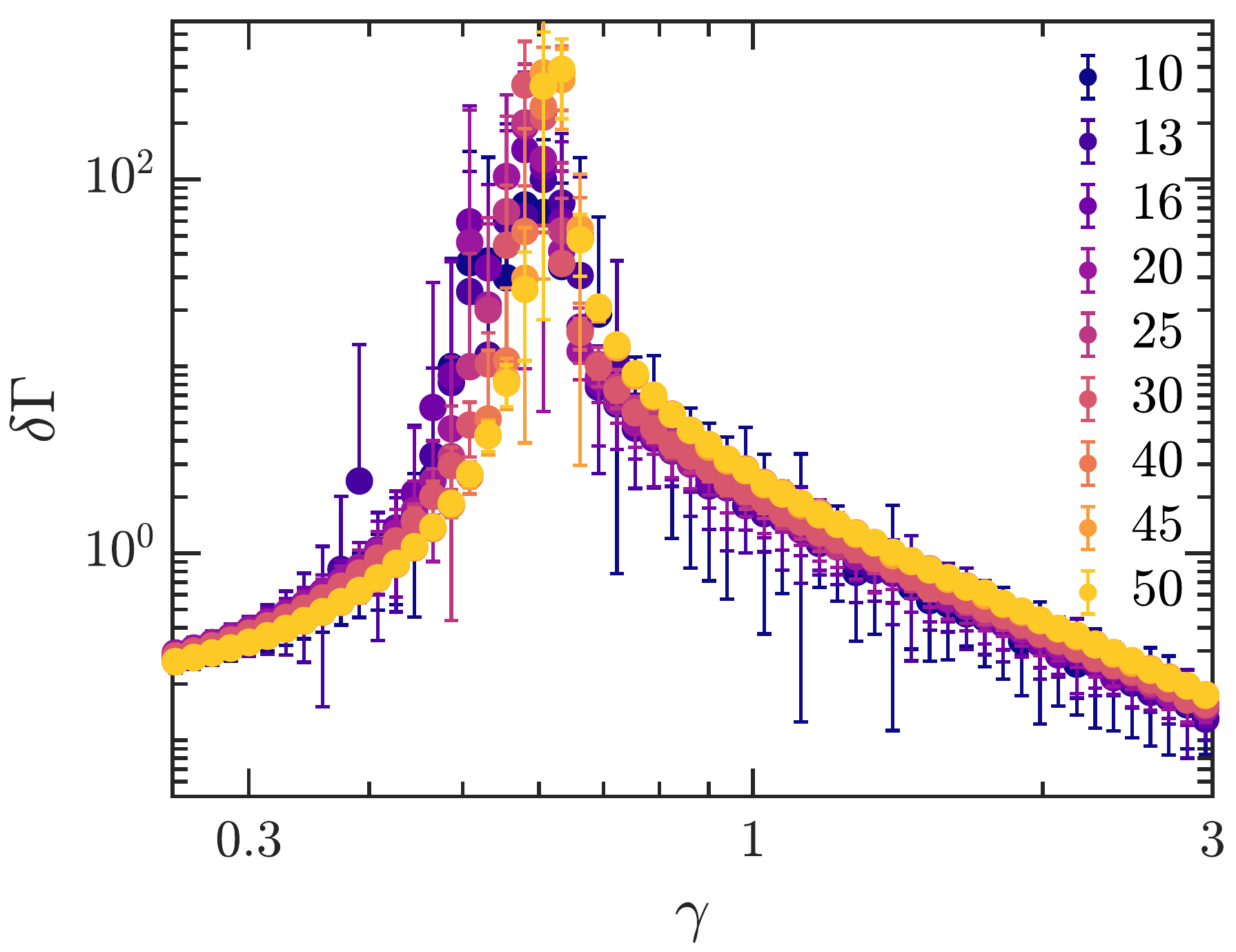}
	\caption{ \label{fig:A_DiffNonaffinity_not_scaled_Z33} The differential nonaffinity parameter versus shear strain for 3D PD networks at $z=3.3$. The finite-size scaling analysis of these data is shown in the main text.}
\end{figure}
\FloatBarrier

\subsection*{3D RGG networks at $z=3.3$}

When subjected to nonlinear strains, network structures with long straight fibers such as lattice-based and RGG models are more likely to show an interesting finite-size effect. At the critical strain, these networks can be rigidified by a small connected cluster of consecutive bonds. This small cluster is responsible for the network stability until the strain becomes large enough to involve a larger portion of network bonds. Therefore the participation ratio of bonds, defined as the ratio of bonds under finite force to all bonds, exhibit a jump between a small value to a larger one. This results in a two-branch behavior in the shear modulus $K$, as can be observed in Fig.\ A.3. As the network size becomes larger, its geometric structure becomes more isotropic, hence this artifact occurs less frequently. However, since we are dealing with 3D systems, it is computationally challenging to simulate large system sizes. As a result, significant number of random samples of our RGG model exhibit this finite-size effect. Note that because of the isotropic nature of PD models, this artifact is rare in those networks.

Since the shear modulus for the samples with a two-branch behavior cannot be fit as a power-law, we remove these samples when calculating the $f$ exponent. Figure\ A.4 b shows the finite-size scaling analysis of $K$ for the RGG model at $z=3.3$ after removing these samples.

\begin{figure}[!h]
	\includegraphics[width=10cm,height=10cm,keepaspectratio]{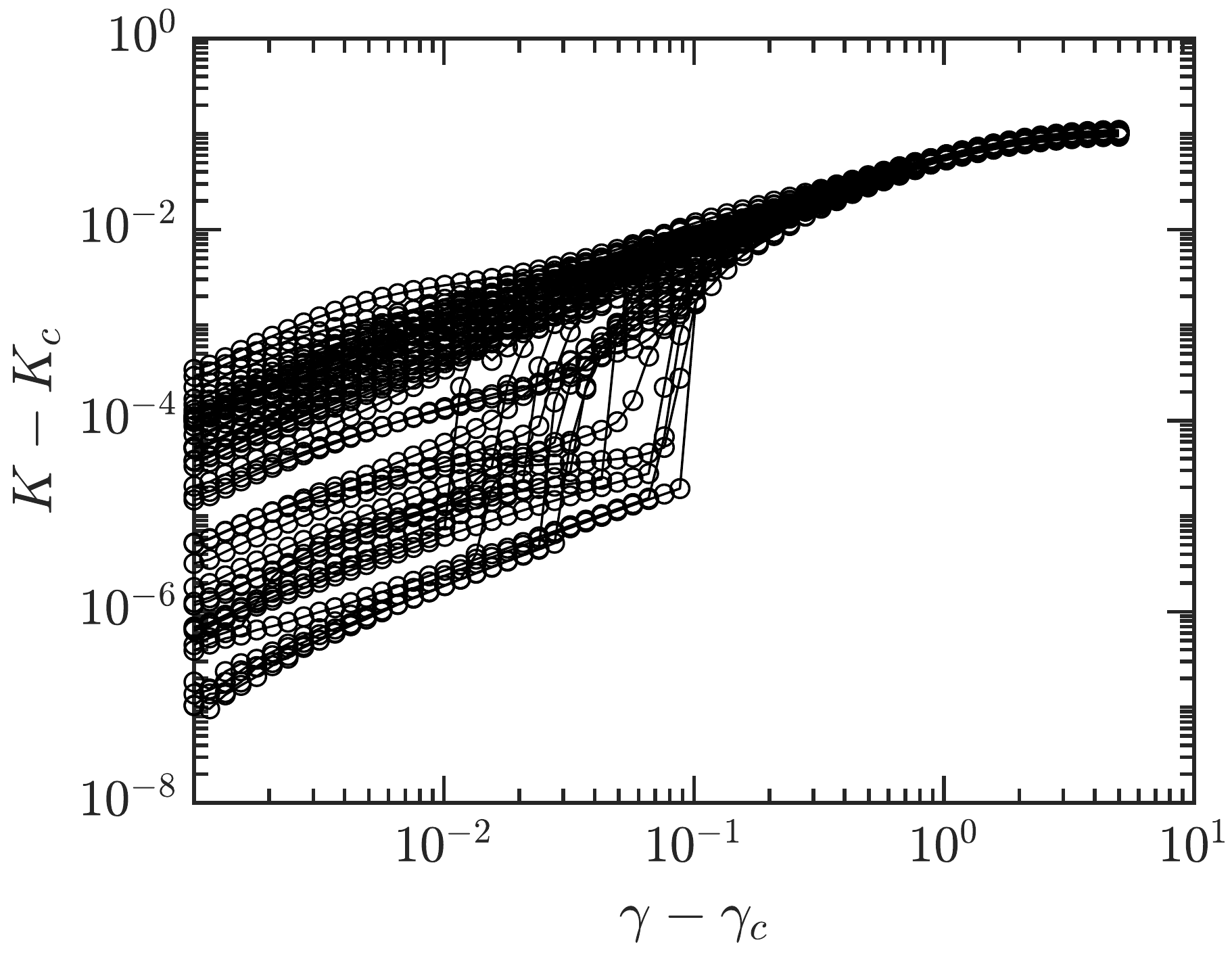}
	\caption{ \label{fig:A_W25_samples_Z33_RGG} The excess shear modulus versus the excess shear strain for 40 random samples of RGG networks at $z=3.3$ and $W=25$. A significant number of samples exhibit a two-branch behavior, an artifact of this specific geometry at small sizes.}
\end{figure}
\FloatBarrier

\begin{figure}[!h]
	\includegraphics[width=15cm,height=15cm,keepaspectratio]{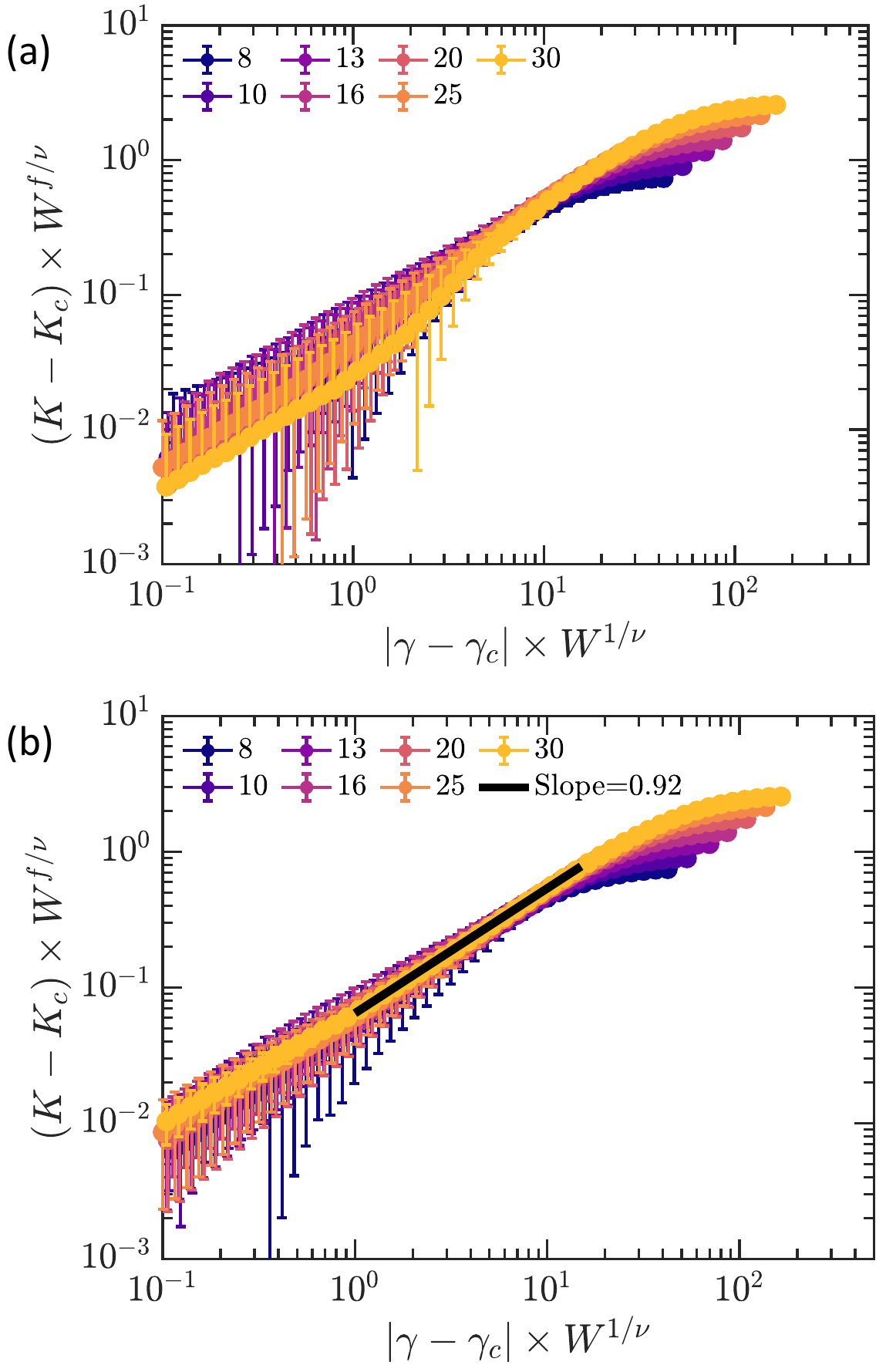}
	\caption{ \label{fig:A_FiniteSizeAnalysis_Z33_RGG} The finite-size scaling analysis of the RGG model at $z=3.3$. (a) Showing the analysis by including all samples. (b) The analysis after removing the random samples that exhibit two-branch behavior. We find an $f = 0.92 \pm 0.02$ by averaging 5 samples of size $W=30$. We used $\nu = (f + 2)/3$.}
\end{figure}
\FloatBarrier

\begin{figure}[!h]
	\includegraphics[width=10cm,height=10cm,keepaspectratio]{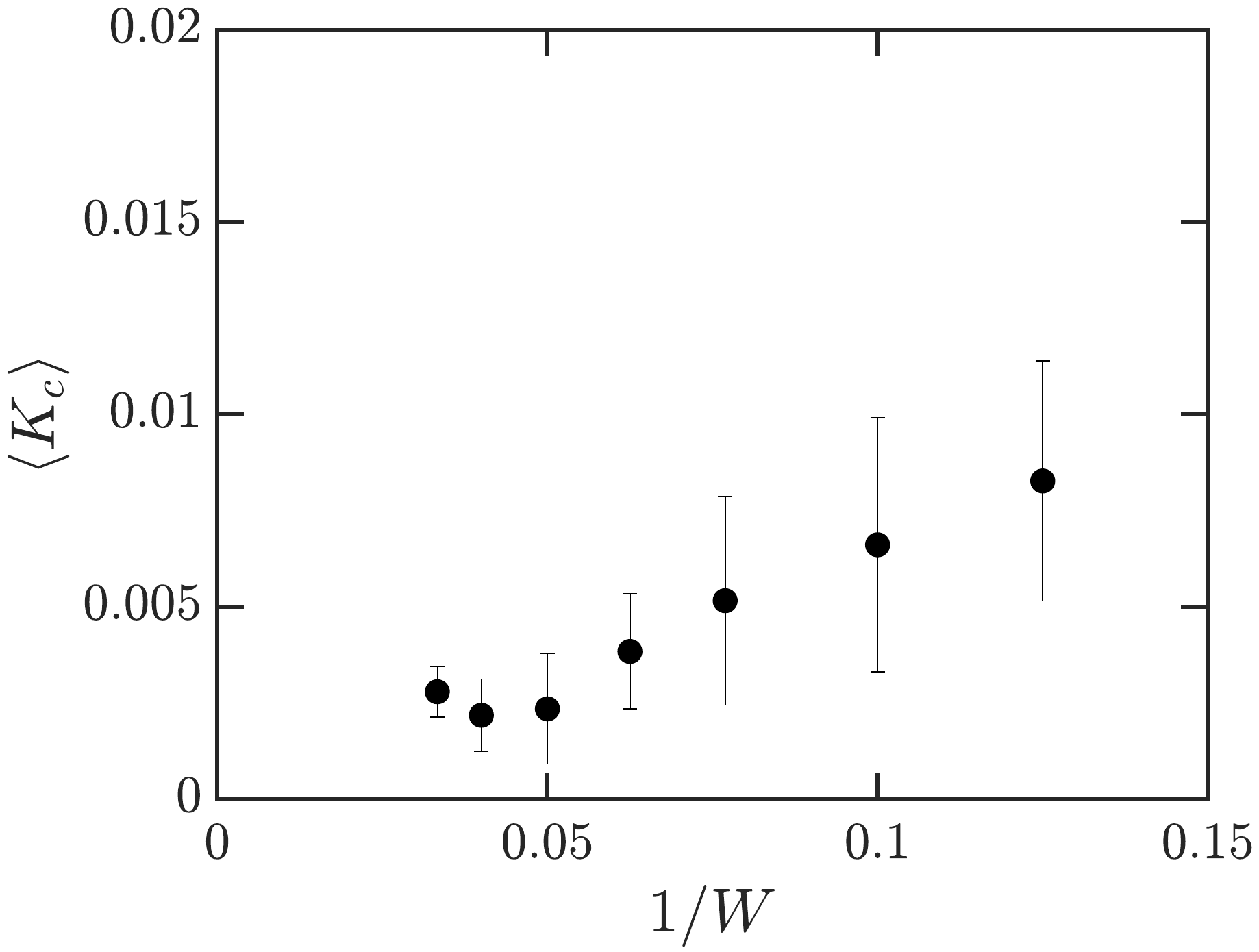}
	\caption{ \label{fig:A_Kc_Z33_RGG} The ensemble average of shear modulus discontinuity $K_c$ for 3D RGG networks at $z=3.3$ versus the inverse of system size $1/W$. Similar to 2D networks, we find that $K_c$ decreases as we increase $W$. However, due to large finite-size effects, our data are not inconsistent with a vanishing $K_c$ in the thermodynamic limit.}
\end{figure}
\FloatBarrier

\begin{figure}[!h]
	\includegraphics[width=10cm,height=10cm,keepaspectratio]{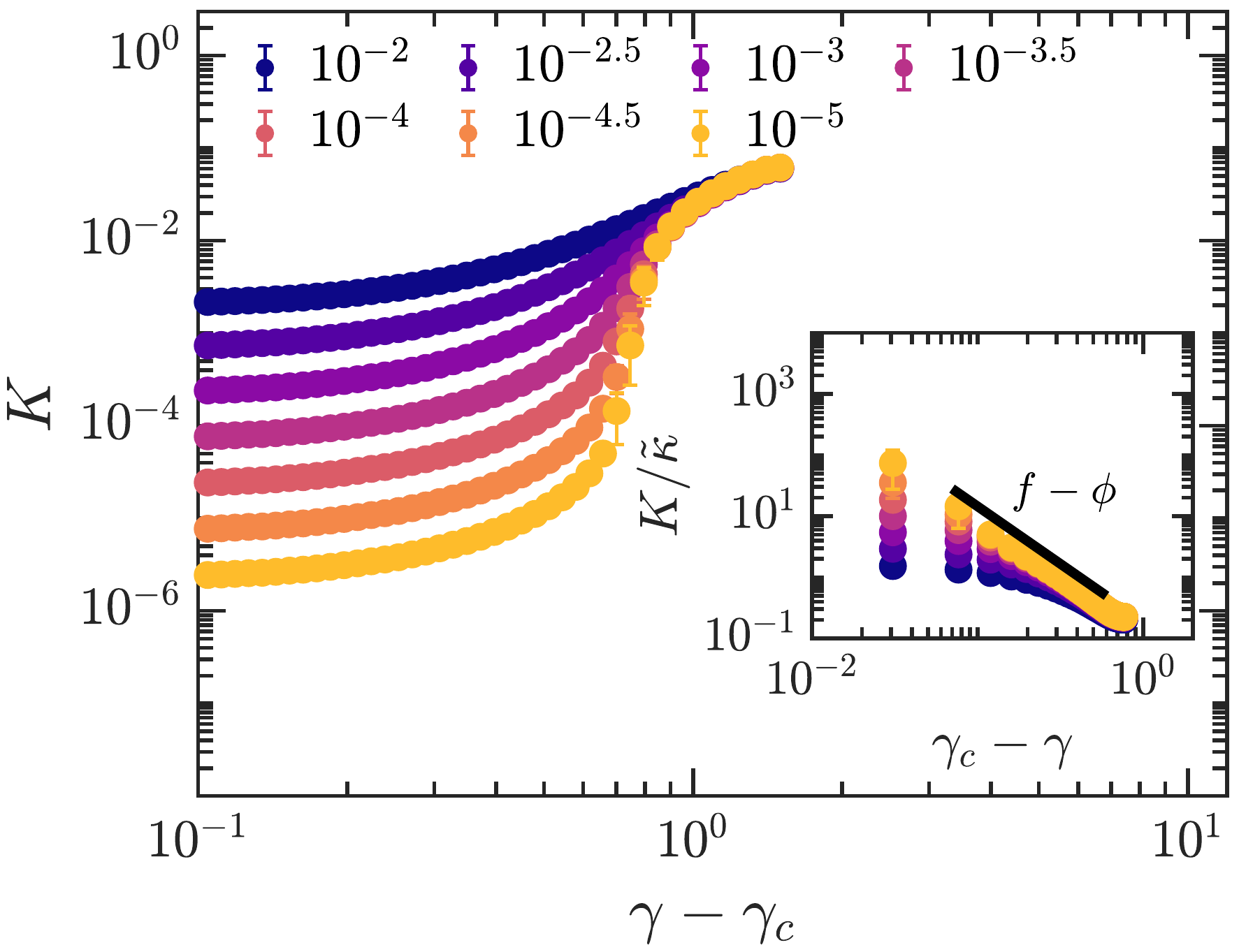}
	\caption{ \label{fig:A_Finite_Kappa_Z33_RGG} Differential shear modulus $K$ for RGG networks with $z=3.3$ and system size $W=30$ at various bending stiffness $\kappa$ as shown in the legend. The inset shows the scaling behavior of $K$ in the subcritical region, where $K \sim \kappa |\gamma - \gamma_c|^{-\lambda}$ with $\lambda = \phi - f$.}
\end{figure}
\FloatBarrier

\begin{figure}[!h]
	\includegraphics[width=10cm,height=10cm,keepaspectratio]{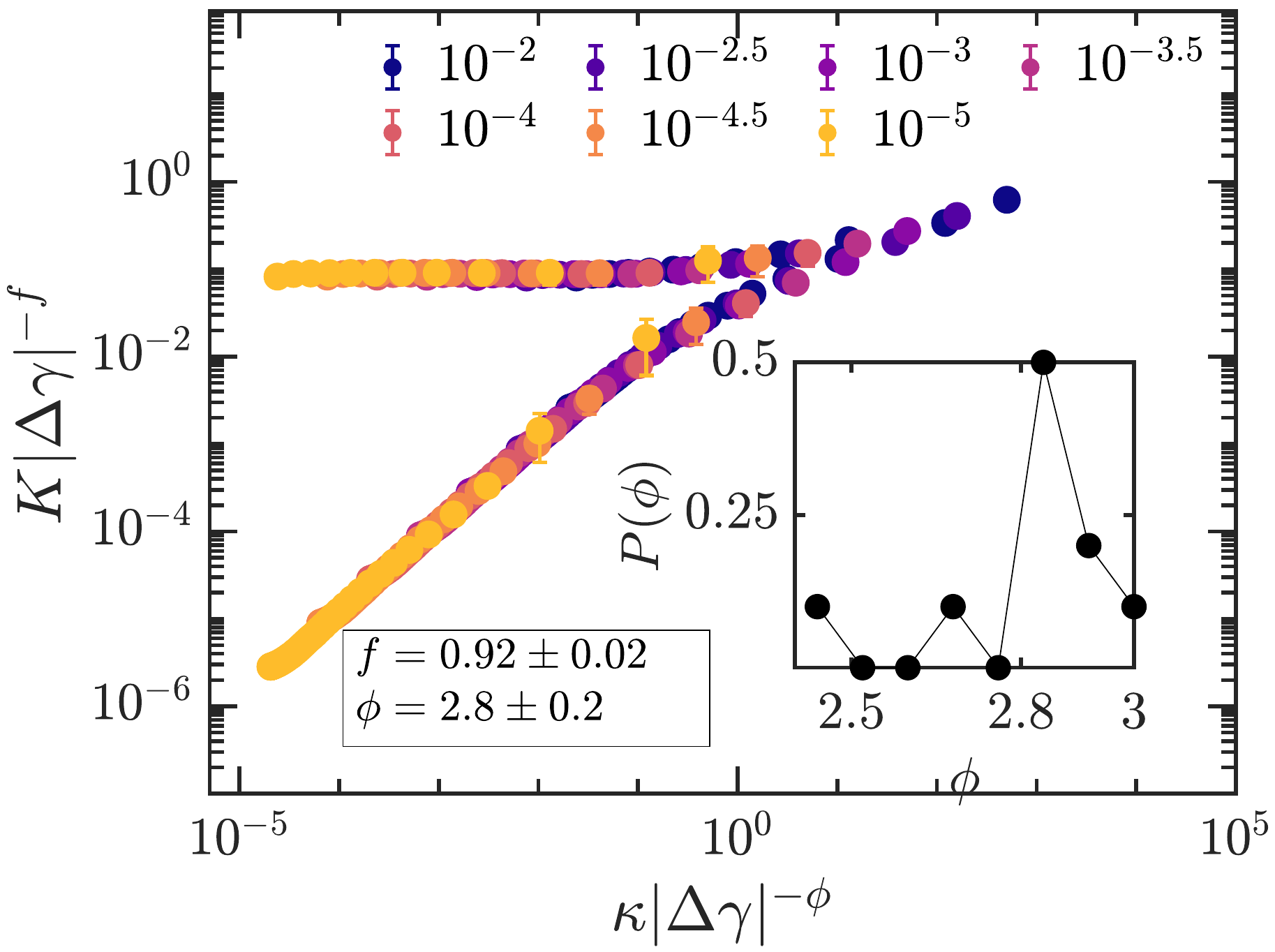}
	\caption{ \label{fig:A_Widom_Z33_RGG}  A Widom-like scaling collapse of differential shear modulus $K$ for data in Fig.\ \ref{fig:A_Finite_Kappa_Z33_RGG}. The critical exponents $f$ and $\phi$ are obtained as explained in the main text. The inset show the distribution of scaling exponent $\phi$.}
\end{figure}
\FloatBarrier

\begin{figure}[!h]
	\includegraphics[width=10cm,height=10cm,keepaspectratio]{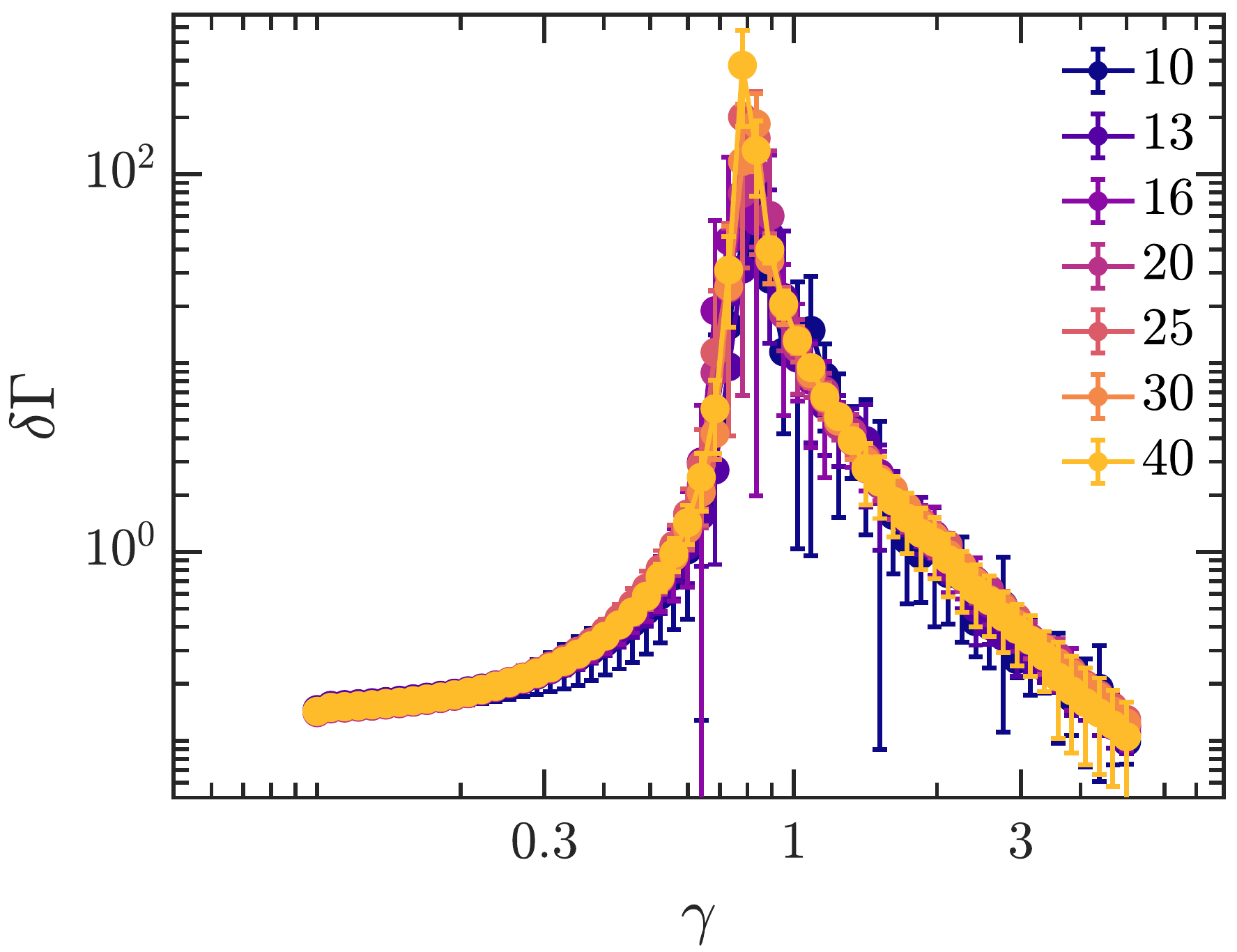}
	\caption{ \label{fig:A_DiffNonaffinity_not_scaled_Z33_RGG} The differential nonaffinity parameter versus shear strain for 3D RGG networks at $z=3.3$. The finite-size scaling analysis of these data is shown in the main text.}
\end{figure}
\FloatBarrier

\subsection*{3D PD networks at $z=4.0$}

The following figures are showing the same analysis that has been performed in the main text for a different network connectivity $z=4.0$. The scaling exponents are close to what we obtained for networks at $z=3.3$. These data again confirm that the scaling relation $f = d \nu -2$ works in 3D. For all of following figures, the data are obtained by averaging 40 random realizations.

\begin{figure}[!h]
	\includegraphics[width=10cm,height=10cm,keepaspectratio]{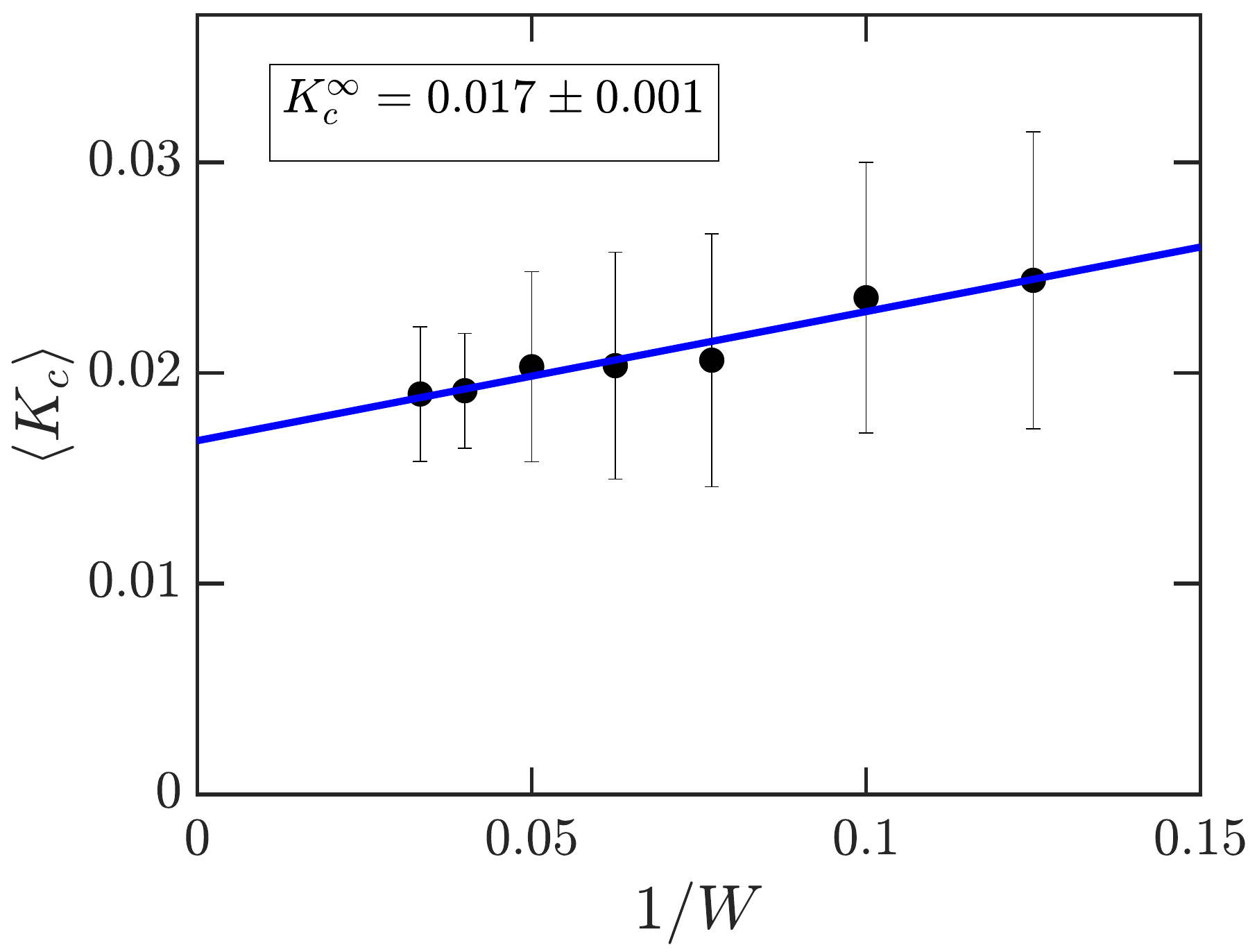}
	\caption{ \label{fig:A_Kc_Z4} The ensemble average of shear modulus discontinuity $K_c$ versus the inverse of system size $1/W$ for 3D PD networks at $z = 4.0$. Similar to 2D networks, we find that $K_c$ decreases as we increase $W$. By fitting a linear equation to the data (shown as a solid blue line), we find a small intercept of 0.017 that is the shear modulus discontinuity in the thermodynamic limit.}
\end{figure}
\FloatBarrier

\begin{figure}[!h]
	\includegraphics[width=10cm,height=10cm,keepaspectratio]{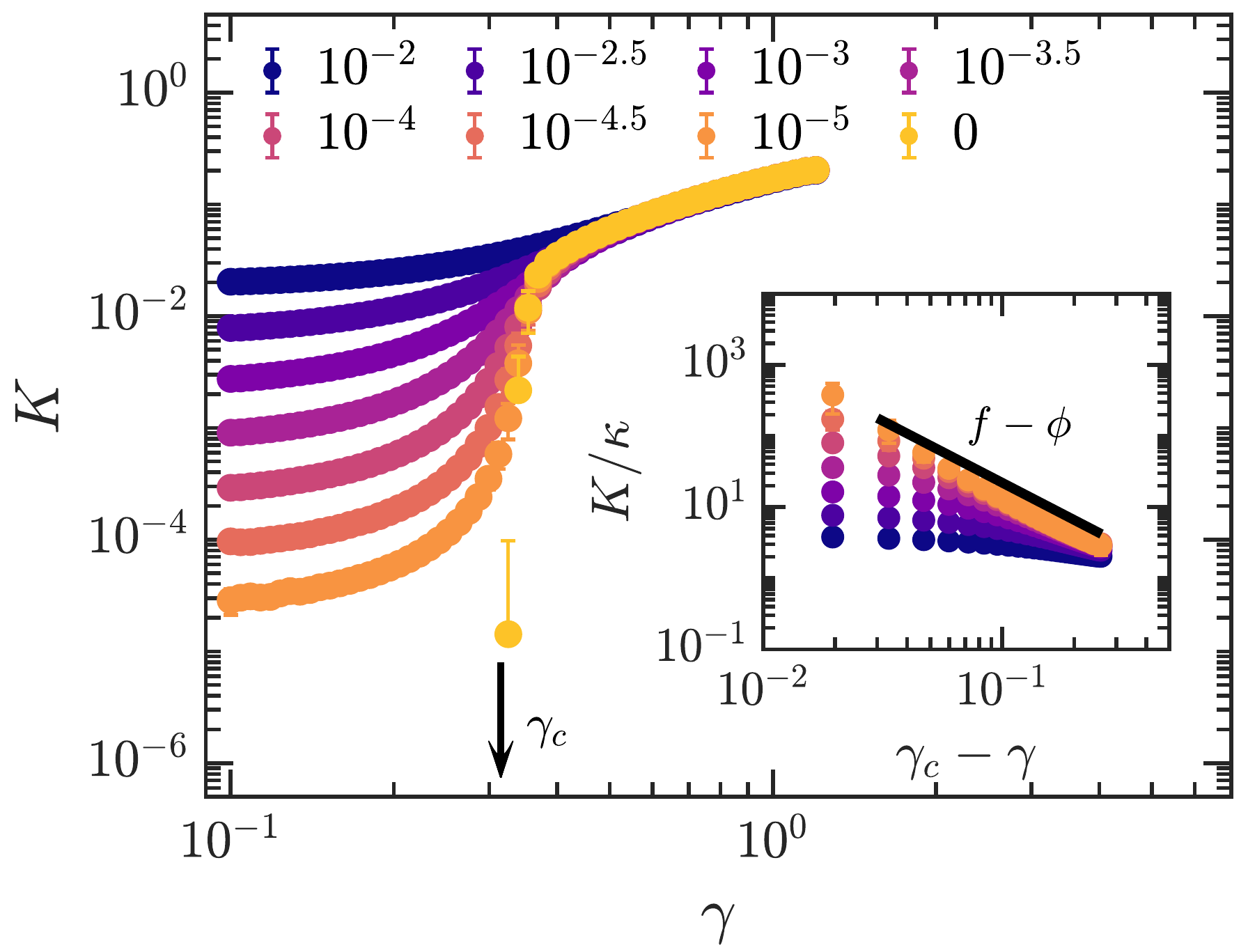}
	\caption{ \label{fig:A_FiniteKappa_Z4} The differential shear modulus of 3D PD networks at $z = 4.0$ with size $W=30$ for different values of bending stiffness between bonds $\kappa$, as shown in the legend. As discussed in the main text, in the subcritical regime $\gamma < \gamma_c$ we expect $K \sim \kappa |\gamma - \gamma_c|^{f - \phi}$. This has been plotted in the inset, $f$ has been already obtained from the finite-size scaling plot of central-force networks and we find $\phi = 2.6 \pm 0.1$ using the data at $\kappa = 10^{-5}$.}
\end{figure}
\FloatBarrier

\begin{figure}[!h]
	\includegraphics[width=10cm,height=10cm,keepaspectratio]{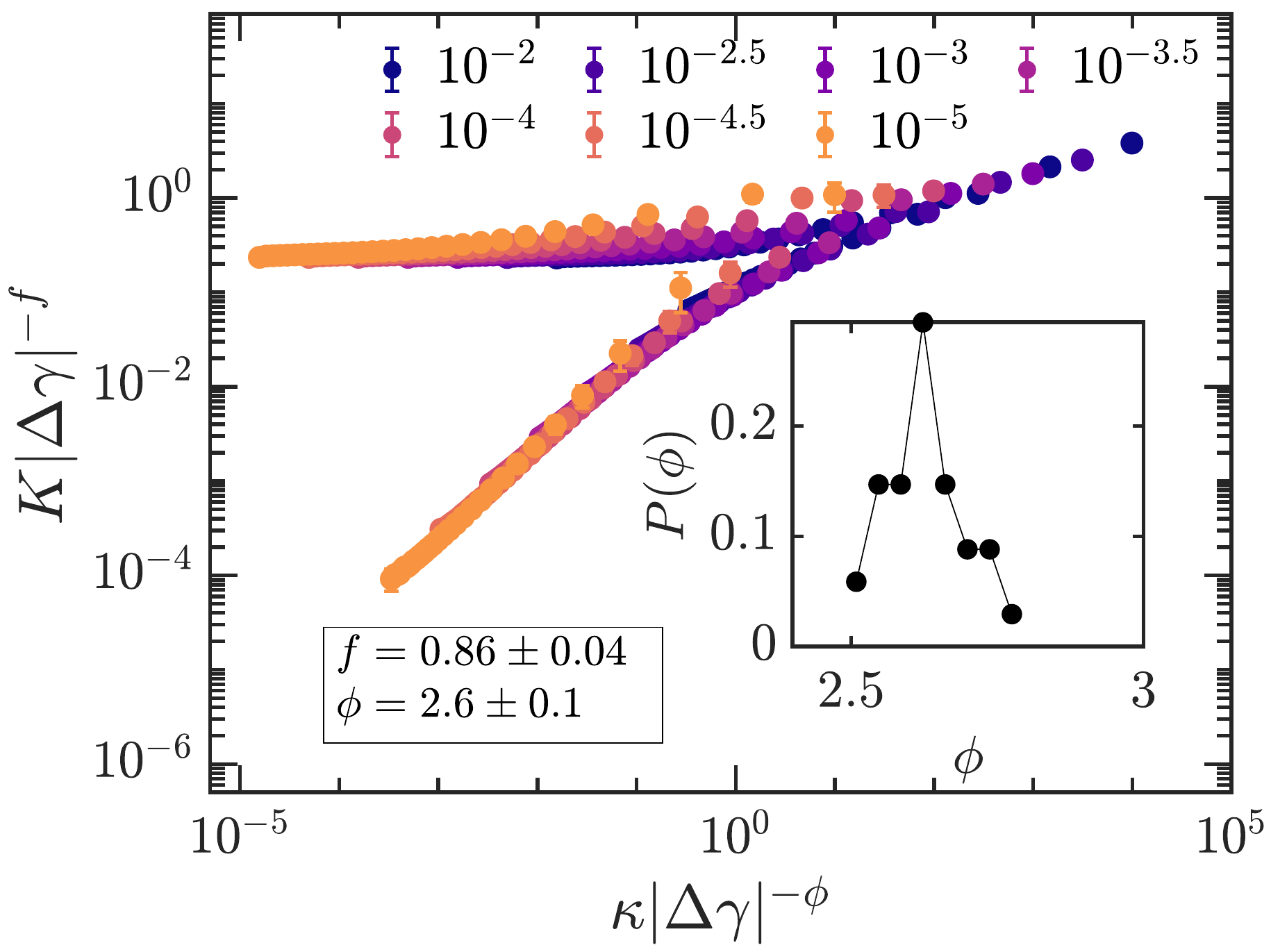}
	\caption{ \label{fig:A_Widom_Z4} A Widom-like scaling collapse of data in Fig.\ \ref{fig:A_FiniteKappa_Z4}. The inset shows the distribution of scaling exponent $\phi$.}
\end{figure}
\FloatBarrier

\begin{figure}[!h]
	\includegraphics[width=10cm,height=10cm,keepaspectratio]{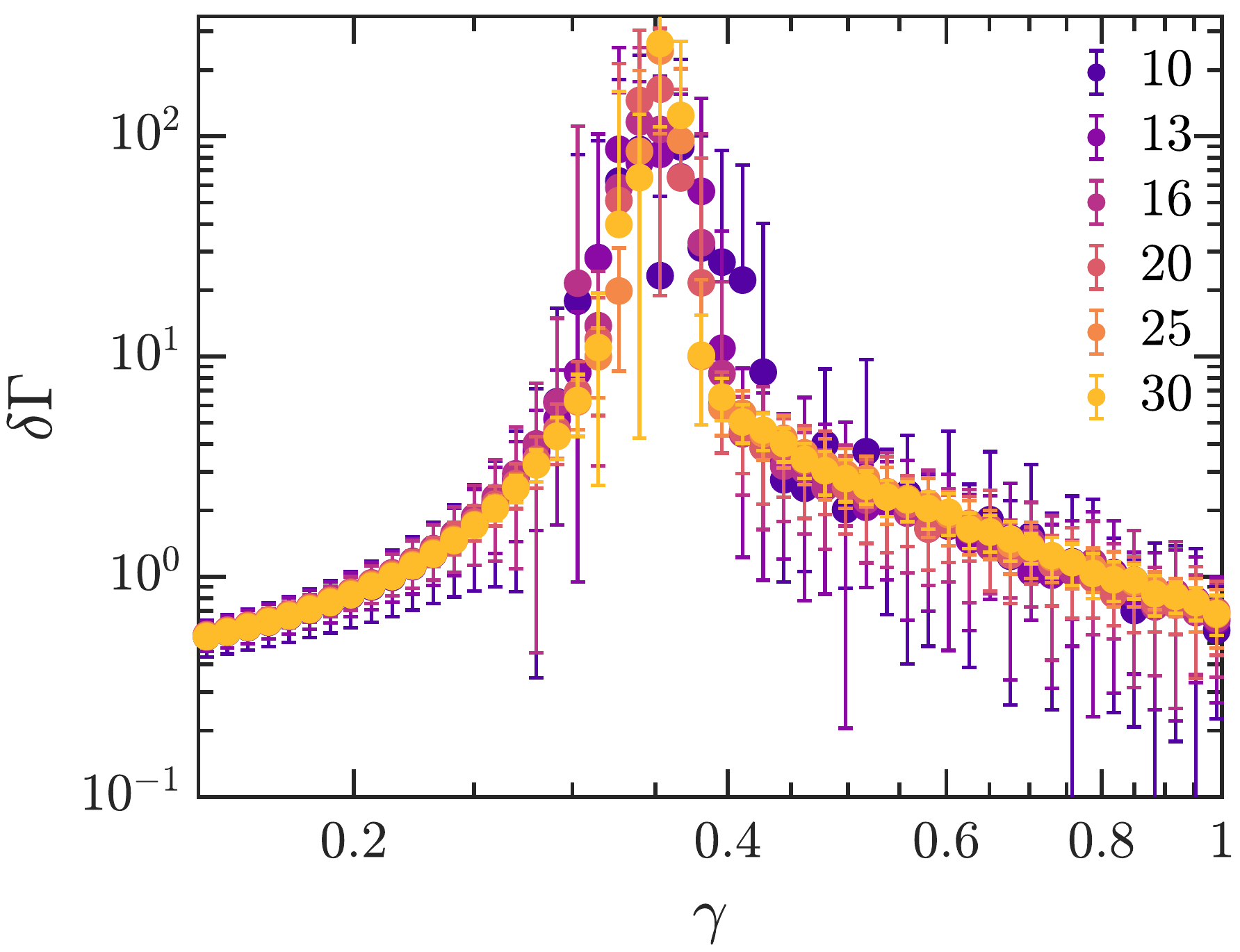}
	\caption{ \label{fig:A_DiffNonaffinity_not_scaled_Z4} The differential nonaffinity parameter for various network sizes as shown in the legend for 3D PD networks at $z = 4.0$.}
\end{figure}
\FloatBarrier

\begin{figure}[!h]
	\includegraphics[width=10cm,height=10cm,keepaspectratio]{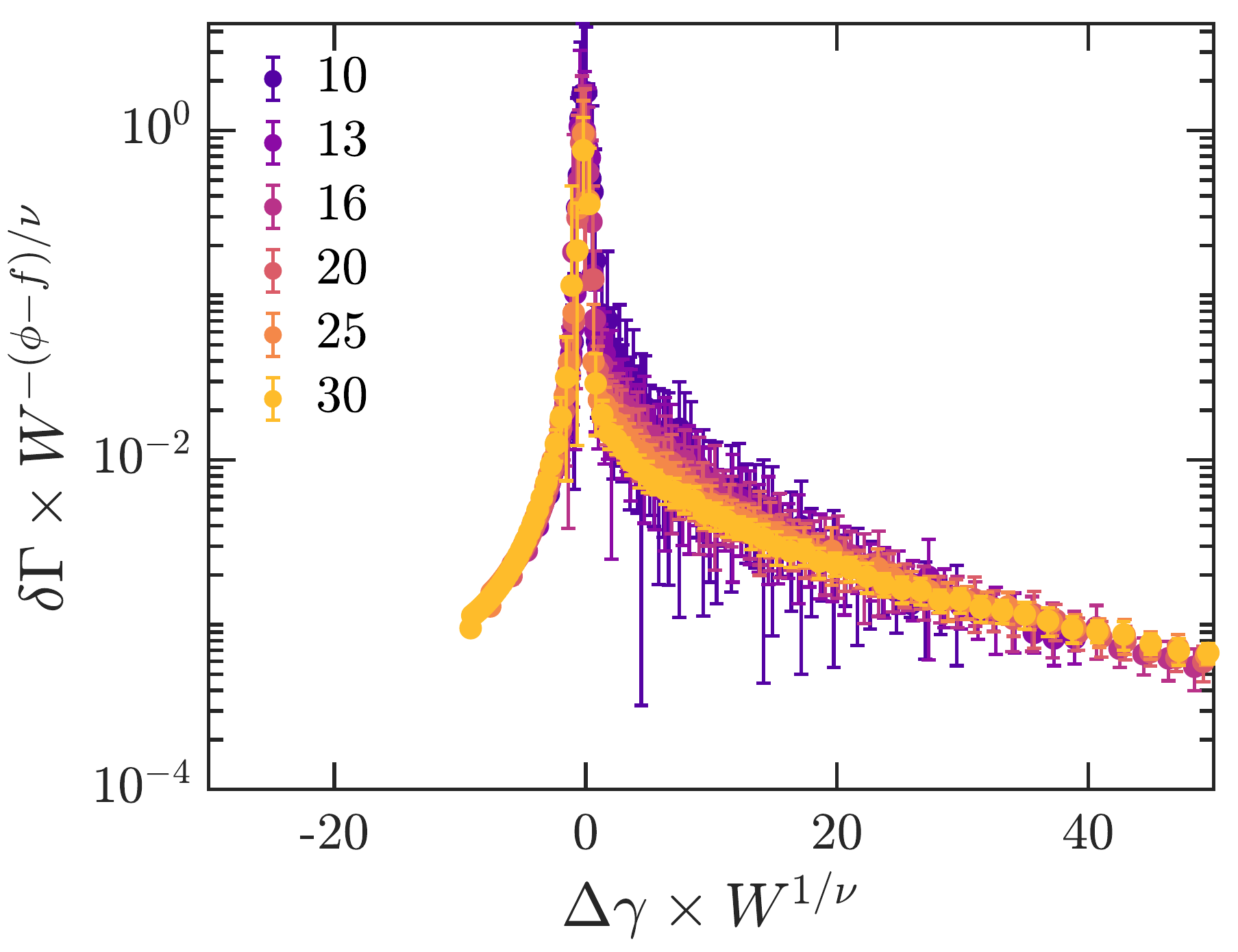}
	\caption{ \label{fig:A_DiffNonaffinity_scaled_Z4} The finite-size scaling collapse of data in Fig.\ \ref{fig:A_DiffNonaffinity_not_scaled_Z4}. The scaling exponents $f$ and $\phi$ have been already obtained. The correlation length exponent $\nu$, however, is calculated using the hyperscaling relation $f = d \nu -2$. This great collapse of nonaffine fluctuations confirms again that the hyperscaling relation holds in 3D networks.}
\end{figure}
\FloatBarrier

\end{document}